\documentclass[twocolumn,showpacs,amsmath,amssymb,prb,showkeys,floatfix]{revtex4}
\usepackage{graphicx}% Include figure files
\usepackage{dcolumn}% Align table columns on decimal point
\usepackage{bm}% bold math

\begin{document}
    
\title{Ferromagnetic resonance study of free hole contribution to magnetization and magnetic anisotropy in 
modulation-doped Ga$_{1 - x}$Mn$_{x}$As/Ga$_{1 - y}$Al$_{y}$As:Be}

\author{X. Liu}
\author{W. L. Lim}
\author{M. Dobrowolska}
\author{J. K. Furdyna}
\affiliation{%
Department of Physics, University of Notre Dame, Notre Dame, IN 46556}
\author{T. Wojtowicz}
\affiliation{%
Department of Physics, University of Notre Dame and Institute of 
Physics, PAS, Warsaw, Poland}
\date{\today}

\begin{abstract}
Ferromagnetic resonance (FMR) is used to study magnetic anisotropy of GaMnAs in a series of 
Ga$_{1 - x}$Mn$_{x}$As/Ga$_{1 - y}$Al$_{y}$As heterostructures modulation-doped by Be. The FMR experiments provide 
a direct measure of cubic and uniaxial magnetic anisotropy fields, and their dependence on the 
doping level.  It is found that the increase in doping -- in addition to rising the Curie temperature 
of the Ga$_{1 - x}$Mn$_{x}$As layers -- also leads to a very significant increase of their uniaxial anisotropy field. 
The FMR measurements further show that the effective $g$-factor of Ga$_{1 - x}$Mn$_{x}$As is also strongly 
affected by the doping. This in turn provides a direct measure of the contribution from the free 
hole magnetization to the magnetization of Ga$_{1 - x}$Mn$_{x}$As system as a whole.
\end{abstract}

%Use showkeys class option if keyword display desired
\maketitle
Magnetic anisotropy in ferromagnetic (FM) semiconductors such as Ga$_{1 - 
x}$Mn$_{x}$As~\cite{Ohno:1998,Ohno:1999,Furdyna:2003} 
is expected to play a key role in future spin-based devices based on these materials.
Although it is now well established that the 
magnetic anisotropy of III$_{1 - x}$Mn$_{x}$V alloys originates from the 
anisotropy of the valence band,~\cite{Abolfath:2001,Dietl:2001,Sawicki:2003} 
the correlation between the magnetic anisotropy and the hole 
concentration is not yet well established. 

It has recently been found that doping the Ga$_{1 - 
y}$Al$_{y}$As barriers of Ga$_{1 - x}$Mn$_{x}$As/Ga$_{1 - y}$Al$_{y}$As 
heterostructures by Be acceptors leads to a significant increase of the Curie temperature 
T$_{C}$ of the Ga$_{1 - x}$Mn$_{x}$As layer.~\cite{Wojtowicz:2003} It was 
also shown that ferromagnetic resonance (FMR) can be used for directly 
determining the magnetic anisotropy of thin FM 
films.~\cite{Farle:1998,Liu:2003} In this paper we use
FMR to show that magnetic anisotropy in ultra-thin 
modulation-doped GaMnAs films changes rapidly with doping level; 
and that the effective $g$-factor of GaMnAs is also strongly affected by the 
doping. This last finding directly reflects the contribution of the free 
hole magnetization to FMR dynamics.

Ferromagnetic Ga$_{1 - x}$Mn$_{x}$As/Ga$_{1 - y}$Al$_{y}$As heterostructures 
were grown on semi-insulating (001) GaAs substrates by molecular beam 
epitaxy (MBE), as described in detail in Ref.~\onlinecite{Wojtowicz:2003}. Three 
heterostructures were used in the current study, all three consisting of a 
5.6 nm Ga$_{1 - x}$Mn$_{x}$As layer ($x$ = 0.062) followed by a 13.5 nm 
Ga$_{0.76}$Al$_{0.24}$As barrier doped with Be starting at the distance of 1 
monolayer away from the Ga$_{1 - x}$Mn$_{x}$As layer. The Be flux was kept constant during the growth, but 
the thickness of the doped region d$_{Be}$ was varied: d$_{Be}$ = 0 (undoped 
control Sample {\#}1), 5.3 nm (Sample {\#}2) and 13.2 nm (Sample {\#}3). The 
FMR measurements were carried out at 9.38 GHz using a Bruker electron 
paramagnetic resonance (EPR) spectrometer. The experimental set-up and the 
polar coordinate system used in the subsequent discussion were described in 
detail in Ref.~\onlinecite{Liu:2003}. Each 
heterostructure was cleaved into three 2mm$\times $2mm square pieces with 
edges along the [110] and $[1 \overline {1}0]$ directions, and the square pieces were 
mounted in the EPR bridge with the $[1 \overline {1}0]$, [110], or [010] directions 
pointing vertically. With the dc magnetic field \textbf{H} in the horizontal 
plane, this allowed us to map out the FMR for \textbf{H} at any angle 
\textit{$\theta $}$_{H}$ between \textbf{H}$\parallel $[001] (normal to the layer 
plane) and three in-plane orientations: \textbf{H}$\parallel $[110], 
$[1 \overline {1}0]$, and [100], following the same procedure as in Ref.~\onlinecite{Liu:2003}.

Because the GaMnAs layers under consideration are extremely thin ($\sim$ 6 
nm), magnetization measurements by SQUID were found to be inaccurate. 
Instead, we made use of the fact that the anomalous Hall 
effect (AHE) is dominated by the magnetization $M$, and can thus serve as a 
measurement of that parameter. To obtain the value of $M$, 
in our analysis we assumed that AHE is dominated by side-jump scattering 
-- i.e., that M $ \propto $ R$_{Hall}$/R$_{sheet}^{2}$, where 
R$_{Hall}$ is the Hall resistance and R$_{sheet}$ is the sheet resistance when 
\textbf{H} is applied perpendicular to the layer.~\cite{Yu:2004} Typical 
magneto-transport data R$_{Hall}$/R$_{sheet}^{2}$ are shown in Fig. 1 for 
several temperatures. Note that at T = 4.22K the magnetization saturates at 
fields of about 5 kOe for undoped material (upper panel), and at higher 
fields (above 6.5 kOe) for modulation-doped samples (bottom panel), 
indicating that the anisotropy field has been modified by the doping.

The inset in Fig. 2 shows FMR spectra at 4.0K for a modulation-doped 
GaMnAs/GaAlAs:Be film (sample {\#}3) in four basic configurations: 
\textbf{H}$\parallel $[001], \textbf{H}$\parallel $[110], 
\textbf{H}$\parallel [1 \overline {1}0]$, and \textbf{H}$\parallel 
$[100]. Strikingly, sharp FMR peaks are observed in all configurations (and 
persist up to T$_{C})$, indicating strong long-range FM coherence of the 
Mn$^{ + + }$ spins. We find this remarkable, since the 5.6nm-thick 
Ga$_{0.94}$Mn$_{0.06}$As film is equivalent to only one monolayer of 
Mn ions randomly distributed over the specimen. As shown in Fig. 2, for 
intermediate orientations of\textbf{ H} between \textbf{H}$\parallel 
$[100] and \textbf{H}$\parallel $[001] the FMR peak $H_{R}$ shifts from 1 
kOe to 10 kOe. Additionally, an EPR peak is observed around 3.3 kOe for all 
field orientations, originating from a small fraction of isolated 
paramagnetic Mn$^{ + + }$ ions with $g$ = 2.00. By their strong dependence 
on crystal geometry, the FMR spectra in Fig. 2 thus establish, that 
magnetic anisotropy plays a major role in determining the fields at which the 
resonances occur.~\cite{Sasaki:2002} 

The magnetic anisotropy parameters of a GaMnAs film can be obtained by 
analyzing the angular dependence of $H_{R}$ using the following equations and 
the coordinate system defined in Ref.~\onlinecite{Liu:2003}. For \textit{$\varphi $}$_{H}$ = 
45$^{\circ}$ [\textbf{H }and \textbf{M }in the $(1\overline {1}0)$ plane],
\begin{subequations}
\label{eq:1}
\begin{eqnarray}
&&({\omega}/{\gamma})^2=[H_{R}\cos({\theta_H-\theta})+(-4{\pi}M+H_{2\perp}+H_{4\perp}/2\nonumber\\  
&&-H_{4\parallel}/4)\cos2\theta+(H_{4\perp}/2+H_{4\parallel}/4)\cos4\theta]\nonumber\\  
&&\times[H_{R}\cos({\theta_H-\theta})+(-4{\pi}M+H_{2\perp}+H_{4\parallel}/2)\cos^2\theta\nonumber\\  
&&+(H_{4\perp}+H_{4\parallel}/2)\cos^4\theta-H_{4\parallel}];
\end{eqnarray}
\noindent
and for \textit{$\varphi $}$_{H}$ = 0$^{\circ}$ [\textbf{H }and \textbf{M }in the plane (010)],
\begin{eqnarray}
&&({\omega}/{\gamma})^2=[H_{R} \cos ( {\theta _H - \theta } ) + ( - 4\pi M + H_{2 \perp } + H_{4 \perp } / 2 \nonumber\\  
&&- H_{4\parallel } / 2 )\cos 2\theta + ( H_{4\perp } / 2 + H_{4\parallel } / 2 )\cos 4\theta  ] \nonumber\\  
&&\times [ H_{R} \cos ( {\theta _H - \theta } ) + ( - 4\pi M + H_{2 \perp } - 2H_{4\parallel }  )\cos ^2\theta \nonumber\\  
&&+ ( H_{4 \perp } + H_{4\parallel }  )\cos ^4\theta + H_{4\parallel }  ].
\end{eqnarray}
\end{subequations}

Here $H_{2 \perp}$ and $H_{4 \perp}$ represent, respectively, the uniaxial and 
the cubic anisotropy fields perpendicular to the film; the anisotropy in the 
film plane is given by the cubic field $H_{4\parallel}$; \textit{$\omega $} is the angular 
microwave frequency; and $\gamma = g\mu _B \hbar ^{ - 1}$ is the 
gyromagnetic ratio, $g$ being the spectroscopic splitting factor and $\hbar $ 
the Planck constant. To simplify the analysis, we have ignored the 
small in-plane uniaxial anisotropy field $H_{2\parallel}$ associated with 
the difference between the $[1 \overline {1} 0]$ and [110] axes.~\cite{Welp:1} 

To determine the parameters appearing in Eq. (1), we first analyze the 
highly-precise values of FMR fields $H_{R}$ for the three high-symmetry 
directions (\textbf{H} parallel to [100], [110], and [001]).~\cite{Liu:2003} An 
independent determination of the $g$-factor and the three anisotropy fields 
$H_{2 \perp}$, $H_{4\parallel}$ and $H_{4 \perp}$ could not be accomplished 
from the analysis of the these values of $H_{R}$ alone without additional 
constraints, i.e., it was possible to find nearly identical fits for several
values of $g$ and anisotropy fields within experimental error. To reduce the number of 
fitting parameters, we have first imposed the value of $g$ = 2.00 that 
corresponds to individual Mn$^{ + + }$ ions. With this constraint, the data 
for Sample {\#}3 at 4 K yield 
unique solutions of $H_{2 \perp}$ = -4319 Oe, $H_{4\parallel}$ = 739 Oe, 
and $H_{4 \perp}$ = -1933 Oe. Using these values in Eq. (1), we then obtain 
the angular variation of $H_{R}$ shown by the dashed line in Fig. 3. The 
dashed curves clearly depart from the data, indicating that the assumption 
of $g$ = 2.00 was not valid.

On the other hand, we note that -- due to the large compressive -- the 
effect of the cubic $H_{4 \perp}$ term is expected to be completely 
overshadowed by $H_{2 \perp}$, and may be neglected. In our second approach 
we will therefore assume $H_{4 \perp}$ = 0, allowing $g$, $H_{4\parallel}$ 
and $H_{2 \perp}$ as fitting parameters. With this approach, the data for 
Sample {\#}3 yield $g$ = 1.80, $H_{4\parallel}$ = 720 Oe and $H_{2 \perp}$ = 
-5887 Oe. Using these values (and $H_{4 \perp}$ = 0) in from Eq. (1), an 
excellent fit to the angular variation of $H_{R}$ is obtained, as shown by 
the solid curve in Fig. 3. We now use the above results as starting 
parameters to carry out a weighted nonlinear least squares fit to FMR 
positions for all values of \textit{$\theta $}$_{H}$, allowing all four parameters ($g$, 
$H_{4\parallel}$, $H_{2 \perp}$, and \textit{H}$_{4 \perp})$ to vary. The results for 
Sample {\#}3 are: $g$ = 1.80 $\pm $ 0.02, $H_{4 \perp}$ = 8 $\pm $ 110 Oe,
$ H_{4\parallel}$ = 735.3 $\pm $ 20 Oe and 
$H_{2 \perp}$ = -5764 $\pm $ 90 Oe. Not that the relation between the three 
anisotropy, $\vert H_{4 \perp}\vert \ll\vert H_{4\parallel}\vert 
\ll\vert H_{2 \perp}\vert $, confirms our assumption that $H_{4 \perp}$ 
can be neglected as a first approximation. Comparing these rigorous results 
with the parameters obtained from $H_{R}$ observed for the three 
high-symmetry orientations (\textbf{H} parallel to [100], [110], and [001]) 
under the assumption that $H_{4 \perp}$ = 0 shows that the two approaches 
lead to very similar values. In our analysis of the data observed as a 
function of temperature we will therefore use the simpler approach. Finally, 
we note that some anisotropy of the $g$-factor is expected in 2D quantum 
structures.~\cite{Peyla:1993,Winkler:2000} However, a fit 
obtained by replacing $g$ with $g$= ($g_{\parallel
}^{2}$sin$^{2}$\textit{$\theta $} + $g_{ \perp}^{2}$cos$^{2}$\textit{$\theta $})$^{1 / 2}$ cannot be 
distinguished from the fit with an isotropic $g$-factor. We have therefore 
accepted an isotropic $g$-factor as an adequate approximation. 

Returning to Fig. 3, it is clear from the inset that this contribution of 
the holes to the $g$-factor is enhanced as hole concentration increases (i.e., 
the fits departs further from the $g$ = 2.00 curves as the doping level 
increases). The fact that \textit{all} our attempts to fit the data consistently lead to 
an effective $g$-factor smaller than 2.00 indicates a finite contribution of 
the orbital magnetic moment to the magnetization. This can be understood as 
follows. The total magnetization of GaMnAs has two components: a 
contribution from Mn$^{ + + }$ ions, with their pure spin moment 
corresponding to $g$ = 2.00; and the free hole contribution, which includes 
both spin and orbital parts. The magnetic moments of the hole spins are 
described by the Luttinger parameter \textit{$\kappa $}. Since the $p$-$d$ 
exchange integral \textit{$\beta $N}$_{0}$ $>$ 0 and \textit{$\kappa $} $>$ 0 
are widely accepted for GaMnAs,~\cite{Dietl:2001} the moments of the 
holes align themselves in the same direction as the moments of the Mn$^{ + + 
}$ ions. In contrast, the orbital part of the hole moment (determined by 
Landau diamagnetic currents) create, through the spin-orbit interaction, a 
magnetization opposite to that of Mn$^{ + + }$ ions.~\cite{Dietl:2001} Thus the 
effective $g$-factor determining the precession of the total 
\textbf{\textit{M}} is a weighted average of $g$-factors for the Mn-ion 
($g_{Mn})$ and for the hole ($g_{h})$, described by the 
expression,~\cite{Rubinstein:2002}
\begin{equation}
\label{eq:2}
g_{eff} = (M_{Mn}+M_{h})/(M_{Mn}/g_{Mn}+ M_{h}/g_{h}).
\end{equation}

Now in the modulation-doped heterostructures, we obtain a significant 
increase of the free hole concentration compared to ``normal'' GaMnAs, which 
automatically enhances their effect.~\cite{Wojtowicz:2003} Owing to the antiferromagnetic $p$-$d$ exchange 
in GaMnAs between Mn$^{ + + }$ and the holes, the value of $g \approx $ 1.80 
can only be achieved by assuming that the magnetization of the free holes \textit{M}$_{h}$ is 
opposite to that of the Mn$^{ + + }$ ions, and $g_{h}$ is positive, as indeed predicted
in Ref.~\onlinecite{Dietl:2001}. With this constraint, Eq. (2) gives \textit{M}$_{h} \approx $ 
-0.15\textit{M}$_{Mn}$, $g_{h} \approx $ 5, thus indicating that Landau 
currents make significant contributions to the magnetization of the free holes, and must therefore 
be taken into account in III-Mn-As materials. 

The $g$-factors and related magnetic properties for T = 10K are listed in Table 
I for all three specimens under investigation. Although the sheet carrier 
densities $p_{s}$-Hall obtained from Hall measurements are not rigorously 
valid due to the AHE contribution, they nevertheless provide a useful 
indication of the \textit{relative} level of the doping.~\cite{Ruzmetov:2004} The values of 
the hole concentration obtained from T$_{C}$ using the mean field 
model,~\cite{Wojtowicz:2003} $p_{v}$-MF, are also listed in Table I for comparison. One 
should note that the concentration of Mn (both substitutional and 
interstial) is the same in all samples, because deposition of the 
modulation-doped GaAlAs:Be layer \textit{after} the GaMnAs layer does not affect the 
GaMnAs layer which was already in 
place.~\cite{Wojtowicz:2003,Yu:2004} The observed changes 
listed in Tabe I for all three samples -- decrease of the $g$-factor with 
doping, enhancement of the uniaxial anisotropy field $H_{2 \perp}$, and 
reduction of $H_{4\parallel}$ -- can thus only be ascribed to changes in 
the hole concentration. 

Measurements of FMR up to the Curie temperature T$_{C}$ enable us to 
determine temperature dependences of both the magnetic anisotropy fields and 
of the $g$-factor in the modulation doped samples. These quantities, obtained 
using the four basic FMR geometries shown in the inset in Fig. 2 and 
assuming $H_{4 \perp}$ = 0, are plotted in Fig. 4 for Samples {\#}1 (undoped; 
open symbols) and {\#}3 (modulation doped; solid symbols). As shown in Fig. 
4(a), FMR occurs above the $g$ = 2.00 resonance position (horizontal 
dash-dotted line) when \textbf{H} is perpendicular to the film, and below 
that position for in-plane \textbf{H} orientations. Shifts from the 
dash-dotted line gradually decrease -- and eventually vanish -- as one 
approaches T$_{C}$. But clearly the modulation-doped sample has a much 
stronger shift (the strongest shift observed in any GaMnAs samples studied 
by FMR) than the undoped sample when \textbf{H} is normal to the film, 
indicating a large increase of magnetic anisotropy due to the doping. 

Figure 4(b) illustrates several basic features of magnetic anisotropy and 
its dependence on temperature and on the free hole concentration. First, we 
note that the cubic anisotropy fields decrease very rapidly with increasing 
T, while $H_{2 \perp}$ drops off much more slowly. And second, modulation 
doping unambiguously increases the perpendicular uniaxial anisotropy field 
$H_{2 \perp}$, while reducing the in-plane cubic field $H_{4\parallel}$. 
These observations are consistent with theoretical calculations predicting 
changes of magnetic anisotropy with hole concentration,~\cite{Dietl:2001} although at 
this point the agreement is only qualitative. Finally, changes of the 
$g$-factor seen in the inset -- although not systematic generally -- show a 
clear trend to decrease in doped samples below $\sim $60K. While these 
$g$-factor values are only approximate, the low-temperature decrease of the 
$g$-factor in the doped samples may reflect the fact that larger numbers of 
hole spins from the GaAlAs barrier will couple with Mn$^{ + + }$ spins in 
GaMnAs as T decreases, thus increasing the effect of Landau diamagnetism on 
the overall magnetization. 

In summary, the results reported in this paper clearly point to the crucial 
role which holes play in determining the magnetic anisotropy and the 
magnetization of Ga$_{1 - x}$Mn$_{x}$As. It is especially important to note 
that magnetic anisotropy of III$_{1 - x}$Mn$_{x}$V materials can be 
manipulated by controlling the hole concentration, thus providing a 
mechanism which may be employed in future device applications. Since the 
dependence of the magnetic anisotropy parameters on the free hole 
concentration is not yet well understood, further rigorous theoretical 
studies addressing this issue are clearly needed. 

This work was supported by the DARPA SpinS Program, by the Director, Office 
of Science, Office of Basic Energy Sciences, and by NSF-NIRT Grant 
DMR02-10519.

\begin{table}
\caption{\label{tab:1}Key parameters for the Ga$_{1 - x}$Mn$_{x}$As/Ga$_{1 - 
y}$Al$_{y}$As heterostructures modulation-doped by Be.}
\begin{ruledtabular}
\begin{tabular}{cccc}
Sample {\#}& 
1& 
2& 
3\\
\hline
Structure& 
GaMnAs/GaAlAs& 
\multicolumn{2}{c} {GaMnAs/GaAlAs:Be} \\
d$_{Be}$ (nm)& 
0& 
5.3& 
13.2\\
T$_{C}$ (K)& 
72& 
85& 
95\\
$p_{v}$-MF (cm$^{ - 3})$& 
1.24$\times $10$^{20}$& 
1.48$\times $10$^{20}$& 
1.64$\times $10$^{20}$\\
$p_{s}$-Hall (cm$^{ - 2})$& 
1.32$\times $10$^{14}$& 
1.74$\times $10$^{14}$& 
2.94$\times $10$^{14}$\\
$M^{10K}$ (emu/cm$^{3})$& 
30.0& 
25.9& 
29.4\\
$H_{2 \perp }^{10K}$ (Oe)& 
-3734& 
-5560& 
-5924\\
$H_{4\parallel }^{10K}$ (Oe)& 
827& 
471& 
557\\
$g^{10K}$& 
1.94& 
1.87& 
1.82\\
\end{tabular}
\end{ruledtabular}
\end{table}

\begin{figure}
\includegraphics[width=1.0\linewidth]{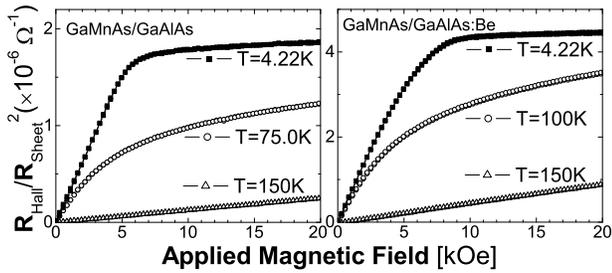}
\caption{Magnetization curves deduced from AHE at various temperatures for 
Samples {\#}1 and {\#}3. Magnetic field \textbf{H} is applied along the hard 
axis of magnetization, \textbf{H}$\parallel $[001].}
\label{Figure1}
\end{figure}

\begin{figure}
\includegraphics[width=1.0\linewidth]{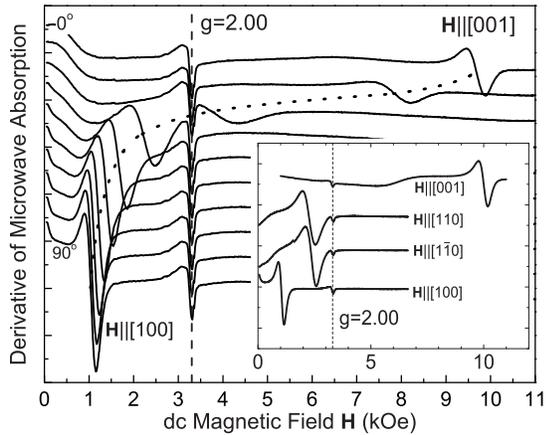}
\caption{FMR spectra (T = 4K) observed for Sample {\#}3 at various 
orientations \textit{$\theta $}$_{H}$ (from 0$^{\circ}$ to 90$^{\circ}$ in 10$^{\circ}$ increments) for 
\textbf{H} between [100] and [001] directions in the (010) plane. The dotted 
line is a guide for eyes, indicating the shifting FMR position. The insert 
shows FMR spectra for the perpendicular (\textbf{H}$\parallel$[001]) 
and three parallel (\textbf{H}$\parallel$[110], \textbf{H}$\vert 
\vert [1\overline {1}0]$, and \textbf{H}$\parallel$[100]) 
configurations observed at 4.0 K. The weak peaks observed at the $g$ = 2.00 
resonance position (indicated by the vertical dashed line) are ascribed to 
EPR of isolated paramagnetic Mn$^{ + + }$ ions.}
\label{Figure2}
\end{figure}

\begin{figure}
\includegraphics[width=1.0\linewidth]{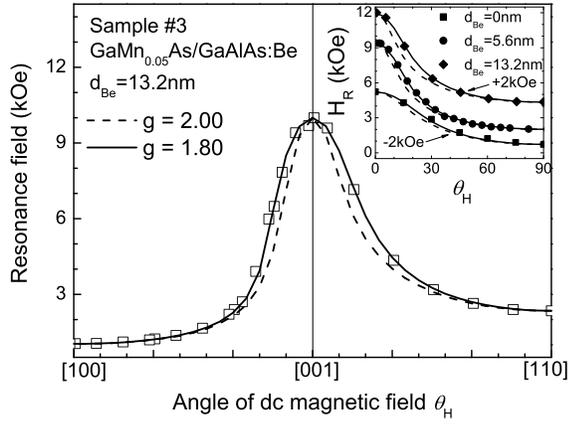}
\caption{Angular dependence of FMR positions for Sample {\#}3 for \textbf{H} 
in the $(1 \overline {1} 
0)$ plane (right-hand panel), and for \textbf{H} in the (010) 
plane (left-hand panel). Inset: Angular dependence of FMR positions at 4.0 K 
for \textbf{H} in the $(1\overline {
1} 0)$ plane for all three samples used in this 
study. Dashed curves show theoretical fits obtained for $g$ = 2.00, $H_{4 \perp
} \ne $ 0. The solid curves are fits obtained for $H_{4 \perp}$ = 0, $g$ = 
1.80 (main figure); and $H_{4 \perp}$ = 0, $g$ =1.80, 1.87, and 1.92 (inset, top 
to bottom).}
\label{Figure3}
\end{figure}

\begin{figure}
\includegraphics[width=1.0\linewidth]{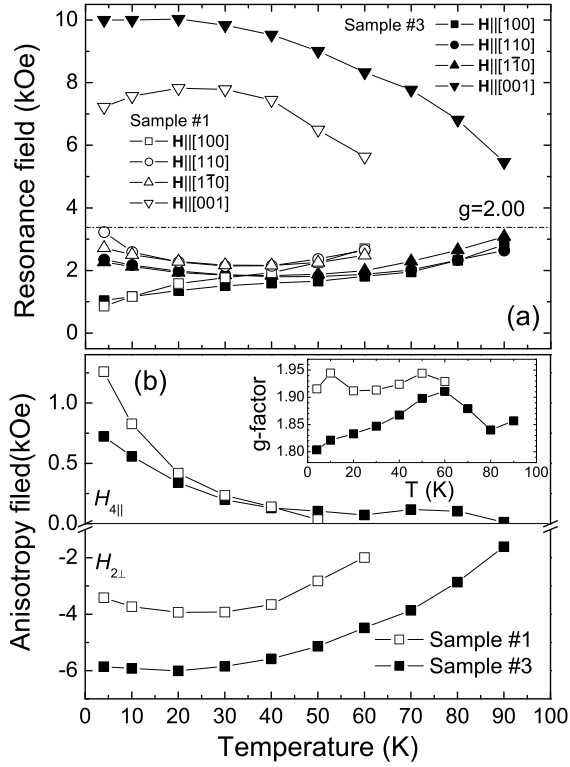}
\caption{Temperature dependence of the FMR results for undoped (Sample {\#}1, 
open symbols) and modulation-doped (Sample {\#}3, solid symbols) 
GaMnAs/GaAlAs heterostructures. Figure 4(a) shows FMR positions observed for 
the four basic orientations of \textbf{H }(same as in the inset in Fig.2). 
Figure 4(b) shows uniaxial anisotropy fields $H_{2 \perp}$ for the two 
samples (bottom panel); cubic anisotropy fields $H_{4\parallel}$ (top 
panel); and the effective $g$-factors (inset).}
\label{Figure4}
\end{figure}

\end{document}